\documentclass[aps,prl,reprint,amsmath,showpacs,superscriptaddress]{revtex4-1}
\usepackage{graphicx}
\usepackage{textcomp}
\usepackage{braket}
\usepackage{color}

\begin{document}
	
	\title{Impact of Many-Body Effects on Landau Levels in Graphene}
	
	\author{J.~Sonntag}
	\affiliation{JARA-FIT and 2nd Institute of Physics, RWTH Aachen University, 52074 Aachen, Germany}
	\affiliation{Peter Gr{\"u}nberg Institute (PGI-9), Forschungszentrum J{\"u}lich, 52425 J{\"u}lich, Germany}
	
	\author{S. Reichardt}
	\affiliation{JARA-FIT and 2nd Institute of Physics, RWTH Aachen University, 52074 Aachen, Germany}
	\affiliation{Physics and Materials Science Research Unit, University of Luxembourg, L-1511 Luxembourg, Luxembourg}
	
	\author{L.~Wirtz}
	\affiliation{Physics and Materials Science Research Unit, University of Luxembourg, L-1511 Luxembourg, Luxembourg}
	
	\author{B.~Beschoten}
	\affiliation{JARA-FIT and 2nd Institute of Physics, RWTH Aachen University, 52074 Aachen, Germany}
	
	\author{M. I. Katsnelson}
	\affiliation{Radboud University, Institute for Molecules and Materials, 6525AJ Nijmegen, Netherlands}
	
	\author{F.~Libisch}
	\affiliation{Institute for Theoretical Physics, Vienna University of Technology, 1040 Vienna, Austria}
	
	\author{C.~Stampfer}
	\affiliation{JARA-FIT and 2nd Institute of Physics, RWTH Aachen University, 52074 Aachen, Germany}
	\affiliation{Peter Gr{\"u}nberg Institute (PGI-9), Forschungszentrum J{\"u}lich, 52425 J{\"u}lich, Germany}
	
	\date{\today}

	\begin{abstract}
		We present magneto-Raman spectroscopy measurements on suspended graphene to investigate the charge carrier density-dependent electron-electron interaction in the presence of Landau levels.
		Utilizing gate-tunable magneto-phonon resonances, we extract the charge carrier density dependence of the Landau level transition energies and the associated effective Fermi velocity $v_\mathrm{F}$.
		In contrast to the logarithmic divergence of $v_\mathrm{F}$ at zero magnetic field, we find a piecewise linear scaling of $v_\mathrm{F}$ as a function of charge carrier density, due to a magnetic field-induced suppression of the long-range Coulomb interaction.
		We quantitatively confirm our experimental findings by performing tight-binding calculations on the level of the Hartree-Fock approximation, which also allow us to estimate an excitonic binding energy of $\approx6\,\mathrm{meV}$ contained in the experimentally extracted Landau level transitions energies.
	\end{abstract}

	\maketitle
	
	\textcolor{black}{
		Many-body interactions crucially influence the electronic properties of graphene~\cite{Kotov2012}.
		They are essential to the understanding of such effects as unconventional quantum Hall states~\cite{Dean2011,Young2012}, graphene plasmonics~\cite{Grigorenko2012,Lundeberg2017,Basov2016} or the formation of a viscous Dirac fermion liquid~\cite{Bandurin2016,Crossno2016}.
		In particular, the long-range electron-electron interaction is predicted to heavily modify the single-particle band structure close to the charge neutrality point (CNP), which is described by a logarithmically divergent effective Fermi velocity $v_\mathrm{F}$~\cite{Gonzalez1994,Gonzalez1999,Stauber2017,DasSarma2013}.
		This charge carrier density ($n_\mathrm{el}$)-dependent band structure renormalization at low or vanishing magnetic fields was experimentally confirmed with various different experimental techniques such as, transport measurements~\cite{Elias2011}, quantum capacitance measurements~\cite{Yu2013}, angle-resolved photoemission spectroscopy (ARPES)~\cite{Bostwick2010, Siegel2011}, and scanning tunneling spectroscopy~\cite{Chae2012,Luican2011}.
		Still, very little is known about the effects of electron-electron interaction on the band structure of graphene in the presence of quantizing magnetic fields, i.e., Landau levels (LLs).
		The only experimental~\cite{Faugeras2015} and theoretical~\cite{Sokolik2017} studies so far focused on the scaling of the effective Fermi velocity with magnetic field at fixed charge carrier density close to the CNP.
		Interestingly, the extracted $v_\mathrm{F}$ is not in agreement with earlier experiments at low magnetic fields~\cite{Elias2011,Yu2013} and hint toward a non-divergent behavior at the CNP.
		This raises the question whether the $n_\mathrm{el}$-dependent renormalization of $v_\mathrm{F}$ and thus the many-body effects are fundamentally different in the presence of LLs.
	}
	
	In this Letter, we report on extracting the renormalized LL energies \textcolor{black}{and the corresponding $v_\mathrm{F}$} as a function of charge carrier density $n_\mathrm{el}$ by optically probing gate-tunable magneto-phonon resonances in suspended graphene.
	Magneto-Raman spectroscopy has successfully been used to probe inter-LL excitations in graphene
	~\cite{Faugeras2015,Faugeras2012,Neumann2015,Yan2010,Faugeras2011,Goler2012,Faugeras2009,Shen2015,Neumann2015a,Ando2007,Goerbig2007,Faugeras2017} and allow the study of the electron-phonon coupling and excitation lifetimes.
	Most importantly, this technique offers a suitable energy scale for measuring the $B$-field-tunable LL transition energies in the form of the Raman G mode phonon energy ($\approx196$~meV).
	Typically, such energy scales characteristic for LLs are difficult to reach by thermally activated transport, which is the method of choice for extracting \textcolor{black}{the energy-band or} $v_\mathrm{F}$ renormalization at negligible $B$-field~\cite{Elias2011}.
	
	\textcolor{black}{
		To compare the renormalization effects at low and high $B$-fields respectively, it is convenient to introduce an effective Fermi velocity $v_\mathrm{F}$ for high magnetic fields, which captures the complete renormalization of the LL energies due to many body-effects~\footnote{Note that in this context the effective Fermi velocity $v_\mathrm{F}=v_{\mathrm{F},n}(B, n_\mathrm{el})$ is no longer linked to its definition as the slope of the energy bands, but rather  directly describes the renormalization of the LL energies.}.
		Thus, the unique square root dependence of the LL spectrum of massless Dirac fermions has to be modified with a renormalized effective Fermi velocity $v_\mathrm{F}$, which now depends on $B$, $n_\mathrm{el}$, and the LL index~$\pm n$. The LL spectrum including many-body interactions thus reads:	$\varepsilon_{\pm n} = \pm v_{\mathrm{F},n}(B, n_\mathrm{el}) \sqrt{2 e \hbar B n}$.}
	Most interestingly, our study of magneto-phonon resonances (MPRs) as a function of $n_\mathrm{el}$ reveals that the effective Fermi velocity does not scale logarithmically with $n_\mathrm{el}$, as it is the case for $B\approx0$\,T, but rather shows a linear and thus finite dependence.
	We attribute this change in behavior to the suppression of the long-range Coulomb interaction for distances much larger than the magnetic length $l_B = \sqrt{\hbar/(e B)}$.
	Moreover, we present a quantitative description of the evolution of $v_\mathrm{F}$ in the presence of LLs within a tight-binding model~\cite{Chizhova2015}
	on the level of the Hartree-Fock approximation, finding a near-perfect agreement with our measurements.

	\begin{figure}
		\includegraphics[width=\columnwidth]{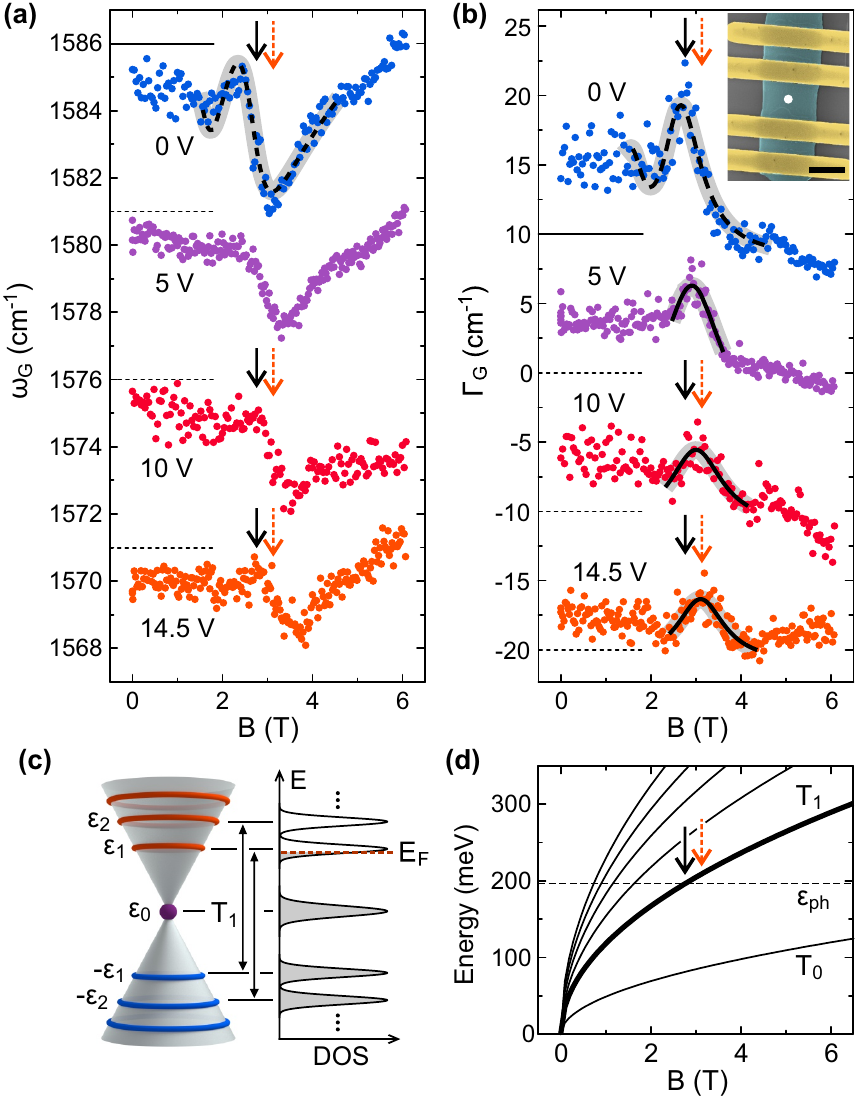}
		\caption{
			The position (a) and width (b) of the Raman G~peak as a function of the magnetic field for different gate voltages.
			For clarity the graphs are each offset by 5\,cm$^{-1}$ and 10\,cm$^{-1}$, respectively (horizontal dashed lines).
			The solid, black and dashed, orange arrows illustrate the shift of the $T_1$-MPR position from
			$V_\mathrm{g} = 0$\,V to $V_\mathrm{g} = 14.5$\,V,
			corresponding to $n_\mathrm{el}\approx0$\,cm$^{-2}$ and $n_\mathrm{el}\approx0.5\times10^{12}$\,cm$^{-2}$, respectively.
			The dashed curves in the upper-most trace represent the theoretical model of MPRs presented in the Supplemental Material~\cite{supp}.
			The solid curves depict Lorentzian fits to $\Gamma_\mathrm{G}$ close to the resonance.
			The inset depicts a false-color scanning electron micrograph of the measured device.
			The white circle portrays our laser spot and the scale bar represents $2$\,\textmu m.
			(c) Schematic illustration of the density of states, the Fermi energy $E_\mathrm{F}$, and the relevant interband LL transition $T_1$.
			(d) Interband LL transition energies $T_n$, assuming an effective $v_\mathrm{F}=1.35{\times}10^6$\,m/s.
			The dashed line indicates the energy of the G mode phonon at $B=0$\,T.}
		\label{fig:MPRs}
	\end{figure}

	For our experimental study, we use a current-annealed suspended graphene device offering high carrier mobility, low intrinsic strain,
	low charge carrier density inhomogeneity, and electron-electron interactions that are not screened by the environment.
	The device consists of a graphene flake on a Si/SiO$_2$ substrate which was exfoliated, contacted with Cr/Au contacts,
	and suspended by etching away $\approx160$\,nm of SiO$_2$.
	A subsequent current annealing step effectively cleans the graphene~\cite{Bolotin2008}, giving rise to a carrier mobility exceeding $400\,000$\,cm$^2$/(Vs)
	and a charge inhomogeneity of less than $n^*{\approx}\, 10^9\,\mathrm{cm^\mathrm{-2}}$ (see Supplemental Material~\cite{supp}),
	which allow the observation of magneto-phonon resonances below 4\,T
	~\cite{Neumann2015,Yan2010,Neumann2015a,Faugeras2011,Goler2012,Faugeras2009,Shen2015,Ando2007,Goerbig2007}.
	The Si back gate moreover permits the controlled tuning of the charge carrier density.
	as well as electrical feedthroughs. This permits combined optical and transport experiments.
	We use linearly polarized laser light with an excitation wavelength of 532\,nm,
	a laser power of 0.5\,mW, and a spot size of $\approx500$\,nm.
	For the detection of the scattered light, we employ a CCD spectrometer with a grating of 1200\,lines/mm.
	All measurements were performed at a temperature of 4.2\,K.
	
	To study magneto-phonon resonances as a function of charge carrier density, we vary $n_\mathrm{el} = \alpha \cdot (V_\mathrm{g}-V_0)$
	by tuning the applied gate voltage $V_\mathrm{g}$, where $V_0 = -0.2$\,V accounts for the residual doping.
	We extract the lever-arm $\alpha=3.15{\times}10^{10}$\,cm$^{-2}$V$^{-1}$ from a Landau fan measurement (see Supplemental Material~\cite{supp}).
	For each specific $V_\mathrm{g}$, we sweep the magnetic field from 0\,T to 6\,T, while recording the Raman spectrum.
	To study the coupling of the electronic system to the Raman-active $E_\mathrm{2g}$~mode, we extract the position $\omega_\mathrm{G}$
	and width $\Gamma_\mathrm{G}$ of the Raman G~peak by fitting a single Lorentzian.
	Their evolution as a function of $B$-field for different values of $V_\mathrm{g}$ is shown in Fig.~\ref{fig:MPRs}a and b, respectively.
	We observe the resonant coupling of the Raman G~mode phonon~\cite{Neumann2015,Neumann2015a,Yan2010,Faugeras2011,Goler2012,Faugeras2009,Shen2015}
	to electronic transitions when its bare energy $\varepsilon_\mathrm{ph} = \hbar \omega_\mathrm{G}(B=0\,\mathrm{T},\,n_\mathrm{el}=0\,\mathrm{cm}^{-2}) \equiv \hbar\omega_0$
	matches the energy of a transition between the discrete LLs.
	Most prominently, this coupling results in a decrease of the phonon lifetime due to the excitation of electron-hole pairs,
	which results in an increased width $\Gamma_\mathrm{G}$ of the Raman G~peak at resonance.
	To first order in perturbation theory, the $E_\mathrm{2g}$-phonon only couples to LL excitations with $\Delta n = \pm 1$~\cite{Ando2007,Goerbig2007}.
	We thus focus on the coupling to LL transitions whose energies are given by $T_n = \varepsilon_{n+1} - \varepsilon_{-n}$ (see Fig.~\ref{fig:MPRs}c and d).
	Note that the influence of excitonic effects on $T_n$ is neglected here and will be discussed later in this Letter.
	The resonance condition $\varepsilon_\mathrm{ph} = T_n$ can be expressed as:
	\begin{equation}
		\label{eq:BTn}
		\hbar\omega_0 = v_{\mathrm{F},T_n}(B_{T_n},n_\mathrm{el})\sqrt{2e\hbar B_{T_n}}(\sqrt{n+1}+\sqrt{n}),
	\end{equation}
	where we defined an effective Fermi velocity $v_{\mathrm{F},T_n}$ of the transition $T_n$ via
	$v_{\mathrm{F},T_n} \equiv T_n/(\sqrt{2e\hbar B}(\sqrt{n+1}+\sqrt{n}))$
	~\footnote{In terms of the renormalized Fermi velocities $v_{\mathrm{F},n+1}$ and $v_{\mathrm{F},-n}$ of the respective Landau levels,
		the effective Fermi velocity of the $T_n$ transition reads: $v_{\mathrm{F},T_n}=(v_{\mathrm{F},n+1}\sqrt{n+1} + v_{\mathrm{F},-n}\sqrt{n})/(\sqrt{n+1}+\sqrt{n}).$}.
	By measuring $\omega_0$ and the value of the $B$-field at which the resonance occurs, $B_{T_n}$,
	one can thus extract the effective Fermi velocity $v_{\mathrm{F},T_n}(B=B_{T_n},n_\mathrm{el})$.
	The experimentally determined $v_{\mathrm{F},T_n}$ evidently contains all corrections from electron-electron interactions.
	Hence the position of the magneto-phonon resonance provides a direct probe of the renormalized transition energy, parametrized by an effective Fermi velocity.
	In particular, we are able to probe the charge carrier density dependence of $v_{\mathrm{F},T_n}$ by varying $V_\mathrm{g}$.

\begin{figure}
		\includegraphics[width=\columnwidth]{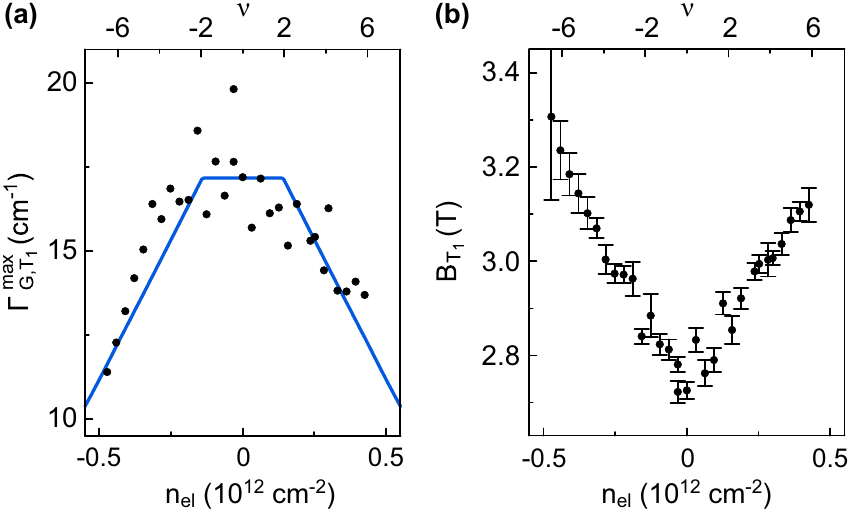}
		\caption{
			(a) Peak width of the G mode at the $T_1$-resonance as a function of charge carrier density.
			The blue line represents the theoretical prediction obtained with the average parameters of all measurements.
			(b) Charge carrier density-dependent position $B_{T_1}$ of the $T_1$-resonance.
			The change in $B_{T_1}$ indicates a change in $v_{\mathrm{F},T_1}$.
			The upper axis shows the filling factor $\nu$ at $B=3$\,T.}
		\label{fig:position}
	\end{figure}

	In the following, we focus on the charge carrier density dependence of the $T_1$ transition~\cite{Neumann2015a,Ando2007,Goerbig2007},
	which gives rise to a resonance at $B_{T_1}\approx3$\,T (Fig.~1a and~b).
	Increasing the charge carrier density leads both to an increase of $B_{T_1}$ (compare black and red arrows in Fig.~\ref{fig:MPRs}a and b)
	and to a decrease of the strength of the $T_1$-MPR, which we define as the maximum value of
	$\Gamma_\mathrm{G}$ at the resonance $B_{T_1}$, $\Gamma_{\mathrm{G},T_1}^\mathrm{max}$.
	For a more quantitative analysis, we fit single Lorentzians to $\Gamma_\mathrm{G}$ (Fig.~\ref{fig:MPRs}b) as a function of $B$ around the $T_1$-MPR
	to obtain $\Gamma_{\mathrm{G},T_1}^\mathrm{\max}$ and $B_{T_1}$ (see Fig.~\ref{fig:position}a and b).
	The observed behavior of $\Gamma_{\mathrm{G},T_1}^\mathrm{max}$ with $n_\mathrm{el}$ can be understood in terms
	of the increasing filling of different LLs and the resulting Pauli blocking.
	For small $|n_\mathrm{el}|$, the Fermi energy stays within the zeroth LL and hence $\Gamma_{\mathrm{G},T_1}^\mathrm{max}$ remains almost constant,
	as the $T_1$ transition involves only the transitions $-1\to+2$ and $-2\to+1$.
	For higher values of $|n_\mathrm{el}|$, the states belonging to the first LL are increasingly filled up and more and more of the degenerate LL-transitions become blocked.
	The decrease of $\Gamma_{\mathrm{G},T_1}^\mathrm{max}$ with $|n_\mathrm{el}|$ is in good agreement with the theoretical prediction
	(blue line in Fig.~\ref{fig:position}a, also see Supplementary Material~\cite{supp}),
	while the linear scaling can be understood from the linear scaling of the filling factor $\nu$ with $n_\mathrm{el}$ (see below).

\begin{figure}
		\includegraphics[width=\columnwidth]{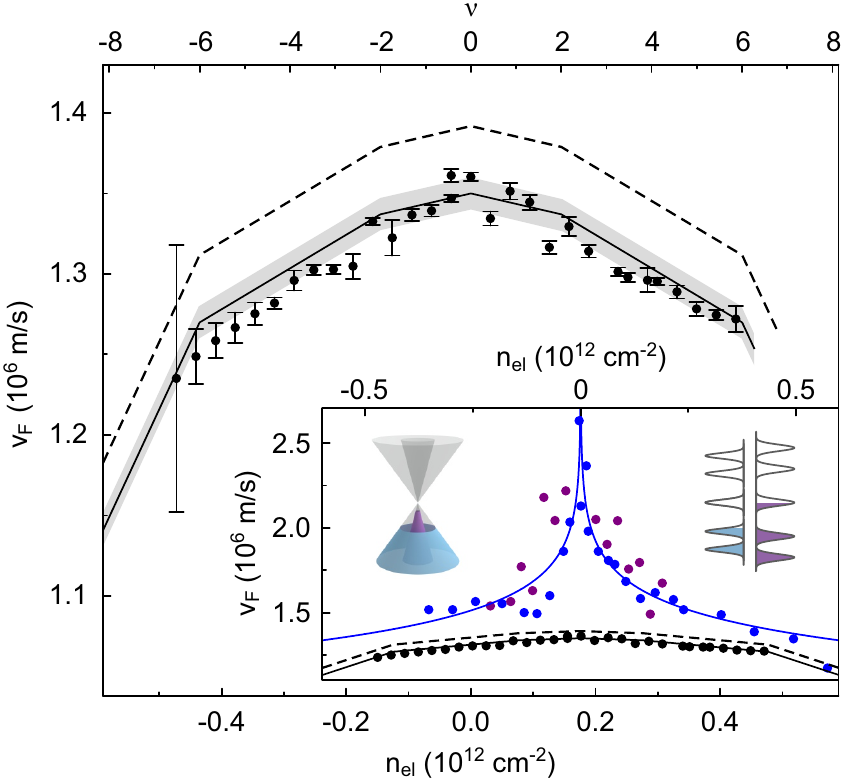}
		\caption{
			The effective Fermi velocity as extracted from the $T_1$-resonance as a function of charge carrier density.
			The upper axis shows the filling factor $\nu$ at $B = 3$\,T.
			The solid black (dashed) line shows the calculated renormalization of $v_{\mathrm{F},T_1}$ without (with) considering the excitonic binding energy.
			The grey-shaded area illustrates the uncertainty due to the $B$\mbox{-}field-dependent renormalization of $v_\mathrm{F}$~\cite{Faugeras2012,Note4}.
			The inset shows a comparison between $v_\mathrm{F}$ in the presence of Landau levels
			(black dots from MPRs, black lines calculated) and at low magnetic field (blue and purple dots).
			The blue data points are taken from Elias~\textit{et al.}~\cite{Elias2011}, while the purple dots represent $v_\mathrm{F}$
			as extracted from temperature-dependent SdHO measurements on the graphene device presented in this study.
			The blue line shows the expected logarithmic renormalization at low magnetic fields.
			The two sketches illustrate the \mbox{$n_{\mathrm{el}}$-induced} renormalization
			of the energy spectrum for low (left) and high (right) magnetic fields.}
		\label{fig:vF}
	\end{figure}

	Next, we analyze the charge carrier density dependence of the position $B_{T_1}$ of the $T_1$-MPR (see Fig.~\ref{fig:position}b).
	According to Eq.~\ref{eq:BTn}, $B_{T_1}$ only depends on the value of the phonon frequency $\omega_0$ and on $v_{\mathrm{F},T_1}$.
	We rule out changes of $\omega_0$ due to tensile strain from electrostatically pulling the graphene flake as the origin of the observed shift in $B_{T_1}$,
	since the observed variation of $\omega_0$ is negligible (less than 2\,cm$^{-1}$, also see Supplemental Material~\cite{supp}).
	Furthermore, tensile strain would \emph{soften} $\omega_0$, i.e., it would lead to a \emph{decrease} of $B_{T_1}$ with increasing $n_\mathrm{el}$.
	Thus, the shift of $B_{T_1}$ can only be caused by a change of the LL excitation energies,
	as described by an $n_\mathrm{el}$-dependent effective Fermi velocity $v_{\mathrm{F},T_1}$.
	For a quantitative analysis of $v_{\mathrm{F},T_1}(n_\mathrm{el})$, we employ Eq.~\ref{eq:BTn} and the extracted $B_{T_1}$
	to directly calculate $v_\mathrm{F}$ as a function of $n_\mathrm{el}$ (see Fig.~\ref{fig:vF}).
	We obtain an effective Fermi velocity ranging from $v_{\mathrm{F},T_1}\approx1.36\times10^6$\,m/s close to the charge neutrality point to $v_{\mathrm{F},T_1}\approx1.24\times10^6$\,m/s at a charge carrier density of $|n_\mathrm{el}|=0.4\times10^{12}$\,cm$^{-2}$.
	Most interestingly, we do not observe a logarithmically divergent behavior close to the CNP,
	as it is the case in the low $B$-field regime (see inset in Fig.~\ref{fig:vF})~\cite{Elias2011,Yu2013}.
	Instead, we find a finite, linear decrease of $v_{\mathrm{F},T_1}$ as a function of $|n_\mathrm{el}|$.
	We attribute this linear behavior to the degeneracy of the states within one LL.
	Due to the degeneracy, the contribution of a certain LL to the renormalization of $v_{\mathrm{F},T_1}$ effectively equals
	the sum of the contributions of all its states weighted by the filling of the LL.
	Since the partial filling factor scales linearly with $n_\mathrm{el}$, so does the renormalization of $v_{\mathrm{F},T_1}$, as long as $n_\mathrm{el}$
	is varied in a range for which the Fermi level $E_\mathrm{F}$ stays within a single LL.
	When $E_{\mathrm{F}}$ enters a different LL, the slope of $v_{\mathrm{F},T_1}$ changes as a different LL is now filled up and its contribution is added.
	
	We confirm this qualitative argument and the experimental observation by quantitative calculations on the level of the Hartree-Fock approximation (HFA)
	within a tight-binding model~\cite{Chizhova2015} (see Supplemental Material~\cite{supp}).
	In the HFA, the single-particle LL energies are renormalized by contributions from all occupied states via the direct Coulomb (Hartree term)
	and exchange interactions (Fock term):
	$\varepsilon_n(B,n_\mathrm{el}) = \varepsilon_n^0(B) + \Sigma^\mathrm{HF}_n(B,n_\mathrm{el})$, where $\varepsilon^0_n$ denotes the bare value of the LL energies and
	\begin{align}
		\label{eq:HFA}
		&\Sigma^\mathrm{HF}_n(B,n_\mathrm{el}) = \frac{1}{N_m} \sum_m \Sigma^\mathrm{HF}_{n,m}(B,n_\mathrm{el}) \\
		&= \frac{1}{N_m} \sum_m  \sum_{n',m'} \bar{\nu}_{n'}(B,n_\mathrm{el})
		\left( 2 v^\mathrm{Hart.}_{\substack{(n,m),\\(n',m')}}(B) - v^\mathrm{Fock}_{\substack{(n,m),\\(n',m')}}(B) \right) \nonumber
	\end{align}
	is the self-energy of LL $n$ in the HFA, averaged over all $N_m$ degenerate states,
	labeled by the quantum number $m$
	~\footnote{The physical meaning of $m$ depends on the choice of gauge for the vector potential. In Landau gauge, $m$ represents the momentum in $y$-direction,
		while in symmetric gauge, $m$ represents the $z$-component of the angular momentum.},
	and $v^\mathrm{Hart.,Fock}_{(n,m),(n',m')}(B)$ represent the direct Coulomb and exchange matrix elements, respectively,
	between the LL states $|n,m\rangle$ and $|n',m'\rangle$.
	Finally, ${\bar{\nu}_n(B,n_\mathrm{el}) = n_\mathrm{el}h/(4eB) - n + 1/2}$ denotes the partial filling factor of LL $n$,
	which is set to 0 (1) for $\bar{\nu}_n<0$ ($>1$) and equals the occupancy of LL $n$.
	Including the Hartree-Fock correction, the energy of the $T_n$-transition reads
	\begin{equation}
		\label{eq:TnwHF}
		T_n(B,n_\mathrm{el}) = \varepsilon^0_{n+1} - \varepsilon^0_{-n} + \Sigma^\mathrm{HF}_{n+1} - \Sigma^\mathrm{HF}_{-n}.
	\end{equation}
	To account for the intrinsic screening of the graphene sheet, we use an effective dielectric constant of $\epsilon=3.1$
	to screen all Coulomb matrix elements by an additional factor of $1/\epsilon$, in agreement with earlier work~\cite{Elias2011}.
	Expressing $T_n(B,n_\mathrm{el})$ in terms of the effective Fermi velocity $v_{\mathrm{F},T_n}$
	(compare Eq.~\ref{eq:BTn}), Eq.~\ref{eq:TnwHF} implies
	\begin{equation}
		\label{eq:vFwHF}
		v_{\mathrm{F},T_n}(n_\mathrm{el}) = v_{\mathrm{F},T_n}(n_\mathrm{el}^0)
		+ \frac{\Delta\Sigma^\mathrm{HF}_{n+1}(n_\mathrm{el}) - \Delta\Sigma^\mathrm{HF}_{-n}(n_\mathrm{el})}{\sqrt{2 e \hbar B_{T_n}}(\sqrt{n+1}+\sqrt{n})},
	\end{equation}
	where $\Delta\Sigma^\mathrm{HF}_{n}(n_\mathrm{el})=\Sigma^\mathrm{HF}_n(n_\mathrm{el})-\Sigma^\mathrm{HF}_n(n_\mathrm{el}^0)$
	denotes the difference in self-energies and $n_\mathrm{el}^0=0$\,cm$^{-2}$.
	Note that in this difference, all contributions from states outside the energy window defined by $n_\mathrm{el}^0$ and $n_{\mathrm{el}}$
	drop out for constant magnetic field, as their occupancies do not change.
	This applies in particular to contributions from states beyond the UV cutoff in renormalization group approaches
	~\cite{Elias2011,Gonzalez1994,Gonzalez1999}, which correspond to contributions from states deep inside the valence band.
	These states only influence the overall scale of $v_\mathrm{F}$, which is set by the value of $v_{\mathrm{F},T_1}(n_\mathrm{el}^0)$.
	For our calculation, we use the experimentally extracted value of $v_{\mathrm{F},T_1}(n_\mathrm{el}^0)=1.35\times10^6\,\mathrm{m/s}$ as input.
	As seen in Fig.~\ref{fig:vF}, \textcolor{black}{our calculation predicts a piecewise-linear $v_{\mathrm{F}}(n_\mathrm{el})$, which is in excellent agreement with our experimental results%
		~\footnote{
			Note that we experimentally extract $v_{\mathrm{F},T_1}$ at slightly different $B$-fields for each value of $n_\mathrm{el}$, due to the $n_\mathrm{el}$-dependent renormalization of $v_{\mathrm{F},T_1}$.
			As shown previously \cite{Faugeras2015}, the effective Fermi velocity also exhibits a $B$-field-dependent renormalization proportional to $\log(B_2/B_1)$ .
			However, the magnitude of this renormalization $\delta v_\mathrm{F}\approx0.02\times10^6$\,m/s for the observed range of $B_{T_1}$-values (2.8\,T to 3.2\,T) is small compared to the $n_\mathrm{el}$-induced corrections and hence we neglect it.}
		and very recent theoretical work~\cite{Sokolik2018}.}

	In order to compare $v_{\mathrm{F},T_1}$ with measurements of the effective Fermi velocity at low magnetic fields extracted by transport experiments~\cite{Elias2011,Yu2013},
	it is important to discuss the so far neglected excitonic effects in our MPR analysis.
	As we probe electron-hole pair excitations, the experimentally extracted LL transition energies
	$T_n=\varepsilon_{n+1}-\varepsilon_{-n}+\varepsilon^\mathrm{bind.}_{n+1,-n}$
	include a (negative) binding energy of the electron-hole pair $\varepsilon^\mathrm{bind.}_{n+1,-n}$.
	Consequently, our experimentally extracted $v_{\mathrm{F},T_1}(n_\mathrm{el}^0)$ already contains an excitonic component of
	$\delta v^\mathrm{bind.}_{\mathrm{F},T_1} = \varepsilon^\mathrm{bind.}_{2,-1}/(\sqrt{2 e \hbar B}(\sqrt{2}+1))$ (compare Eq.~\ref{eq:BTn}).
	To correct for the excitonic effects and thus permit a sensible comparison to the $v_{\mathrm{F}}$ extracted from transport measurements,
	we estimate $\varepsilon^\mathrm{bind.}_{n+1,-n}$ by approximating it within our HFA tight-binding model
	as the difference of the direct and exchange Coulomb matrix elements, averaged over all possible pairs of degenerate LL states:
	\begin{equation}
		\label{eq:epsbind}
		\varepsilon^\mathrm{bind.}_{n+1,-n} = 1/N_m^2 \sum_{m,m'}
		\left( v^\mathrm{Hart.}_{\substack{(n+1,m),\\(-n,m')}} - v^\mathrm{Fock}_{\substack{(n+1,m),\\(-n,m')}} \right).
	\end{equation}
	The numerical evaluation of this expression yields an $n_\mathrm{el}$\mbox{-}independent estimate of $\varepsilon^\mathrm{bind.}_{2,-1}\approx-6$\,meV,
	when including the screening factor of $1/\epsilon$.
	When correcting $v_{\mathrm{F},T_1}$ for $\delta v^\mathrm{bind.}_{\mathrm{F},T_1}$, we obtain values for the effective Fermi velocity without any excitonic effects,
	as shown as the black dotted line in Fig.~\ref{fig:vF}.
	
	The inset in Fig.~\ref{fig:vF} shows a comparison of $v_\mathrm{F}$ at low magnetic fields ($<0.5$\,T) and in the presence of LLs ($\approx3$\,T).
	The purple dots represent $v_\mathrm{F}(n_\mathrm{el})$ at low magnetic fields, extracted from temperature-dependent Shubnikov-de Haas oscillation (SdHO) measurements
	(see Supplemental Material~\cite{supp}) taken on the very same device used for our MPR study.
	They are in good agreement with the previously reported $v_\mathrm{F}$ by Elias~\textit{et al.}~\cite{Elias2011} (blue dots)
	and the expected logarithmic behavior at low magnetic fields (blue line).
	Most interestingly, there is a striking difference in the $n_\mathrm{el}$-dependence between $v_\mathrm{F}$ extracted at low magnetic fields from transport experiments
	and $v_{\mathrm{F},T_1}$ determined at high magnetic fields from our optical measurements.
	Note that the magneto-Raman measurements always probe $v_\mathrm{F}$ away from the Dirac point at approximately half the phonon energy \textcolor{black}{($\approx100\,\mathrm{meV}$)}, while transport experiments extract the band slope at the Fermi surface.
	\textcolor{black}{However, previous ARPES~\cite{Siegel2011} and scanning tunneling spectroscopy~\cite{Chae2012} studies established that the renormalized bands remain linear within an energy window around the CNP of at least 200~meV (see left schematic in inset Fig.~\ref{fig:vF}).	
		As we both probe $v_\mathrm{F}$ and tune the Fermi energy within this energy window, the exact energy at which $v_\mathrm{F}$ is probed is irrelevant.
		Consequently, it is justified to compare our results to the ones from transport measurements at low $B$-fields.}
	Since the excitonic correction cannot account for the change in effective Fermi velocities \textcolor{black}{between the two techniques}, we conclude that the difference in $v_\mathrm{F}$ is \emph{not} due to the way in which $v_\mathrm{F}$ is determined, but rather due to the difference in electron-electron interaction at low $B$\mbox{-}fields and in the presence of LLs.
	At low magnetic fields, the self-energy correction to $v_\mathrm{F}$ diverges due to the long-range behavior of the Coulomb interaction
	and the delocalized nature of the Dirac electrons at the K~point.
	By contrast, high magnetic fields exponentially localize the electronic wave functions once LLs are present,
	with a decay constant on the order of the magnetic length $l_B$ (see Supplementary Material~\cite{supp}).
	As a result, the long-range divergence is eliminated.

	In conclusion, we extracted the charge carrier density dependence of the effective Fermi velocity
	close to the charge neutrality point in the presence of LLs by studying magneto-phonon resonances.
	In contrast to the logarithmic renormalization of $v_\mathrm{F}$ found at low magnetic fields, we find that in the LL regime $v_\mathrm{F}$
	stays finite and scales piecewise linearly with $n_\mathrm{el}$.
	The linear scaling of $v_\mathrm{F}$ with $n_\mathrm{el}$ originates from the degeneracy of the LLs.
	By contrast, the suppression of the divergence at the charge neutrality point can be traced back to
	the spatial confinement of the electron wave functions by the magnetic field, which cuts off the divergent long-range Fock contribution.
	Tight-binding calculations based on the Hartree-Fock approximation quantitatively verify our experimental findings
	and confirm that electron-electron interactions in graphene are indeed very sensitive to the applied magnetic field
	and that they can change dramatically for different magnetic field regimes.
	The general nature of the gained insight into many-body effects on the electronic excitation energies in strong magnetic fields
	make them applicable to the study of other low-dimensional materials as well and can be of great value for the effective tuning of material properties.

	\begin{acknowledgments}
		
		We thank C. Neumann, E. Andrei, and F.~Guinea for helpful discussions and M. Goldsche for help with sample fabrication.
		Support by the ERC (GA-Nr. 280140), the Helmholtz Nano Facility~\cite{Albrecht2017}, the DFG, and the European Union’s Horizon 2020 programme under grant agreement No 696656 (Graphene Flagship) are gratefully acknowledged.
		S.R. and L.W. acknowledge financial support by the National Research Fund (FNR) Luxembourg
		(projects RAMGRASEA and \mbox{TMD-nano}: INTER/ANR/13/20/NANOTMD).
		F.L. acknowledges financial support by the FWF (SFB-F41 ViCoM) and numerical support by the Vienna Scientific Cluster (VSC).
		
	\end{acknowledgments}

%


\widetext
\pagebreak
\begin{center}
	\textbf{\large Supplemental Material: Impact of Many-Body Effects on Landau Levels in Graphene}
\end{center}
\setcounter{equation}{0}
\setcounter{figure}{0}
\setcounter{table}{0}
\setcounter{page}{1}
\makeatletter
\renewcommand{\theequation}{S\arabic{equation}}
\renewcommand{\thefigure}{S\arabic{figure}}
\renewcommand{\thesection}{S\arabic{section}}
\renewcommand{\thetable}{S\arabic{table}}

	\section{Theory of Magneto-Phonon Resonances in Single-layer Graphene}

	\label{sec:MPRTheory}
	Following Ando \cite{Ando2007}, Goerbig~\textit{et al.} \cite{Goerbig2007}, and Neumann~\textit{et al.} \cite{Neumann2015a}, we calculate the Raman G mode phonon frequency $\omega_\mathrm{G}$ and line width $\Gamma_\mathrm{G}$ as the real and doubled negative imaginary part, respectively, of the root of the following equation:
	\begin{equation} \label{eq:MPRSLG}
		\omega^{2}-(\omega_{\mathrm{0}}-i\gamma_{\mathrm{0}}/2)^2  =  2(\omega_{\mathrm{0}}-i\gamma_{\mathrm{0}}/2) \Pi(\omega),
	\end{equation}
	where the phonon self-energy $\Pi(\omega)$ is given by
	\begin{equation} \label{eq:SelfEnergy}
		\Pi(\omega) =  ev_\mathrm{F}^2B\lambda \sum_{n=0}^{\infty}\left[\frac{\left(\bar{\nu}_{- n}-\bar{\nu}_{n+1}-\bar{\nu}_{n}+\bar{\nu}_{-(n+1)}\right)(T_{n}-i\gamma_{\mathrm{el}}/2)}{\omega^{2}-(T_{n}-i\gamma_{\mathrm{el}}/2)^{2}}+\frac{2}{T_{n}-i\gamma_{\mathrm{el}}/2}\right].
	\end{equation}
	Here, $T_{n}=\frac{1}{\hbar}\left|\varepsilon_{n+1}+\varepsilon_{n}\right|$ are the frequencies associated with the inter-Landau level transitions, and $\varepsilon_{\pm n} = \pm v_{\mathrm{F},n}\sqrt{2 e \hbar B n}$ is the energy of the $\pm n$th Landau level as stated in the main text.
	$\bar{\nu}_{n}=(\nu-4n+2)/4$ denotes the partial filling factor, which depends on the filling factor $\nu=n_{\mathrm{el}}h/(eB)$ and obeys $0\leq\bar{\nu}_{n}\leq1$. $\gamma_{\mathrm{el}}$ introduces a damping of the Landau level excitations to account for their finite lifetimes and $\gamma_{\mathrm{0}}$ represents the damping of the phonon mode due to anharmonic effects.
	The dimensionless electron-phonon coupling constant is denoted by $\lambda$.
	By fitting the root of Eq.~\ref{eq:MPRSLG} to all our MPR measurements we get a set of average parameters leading to: $\omega_{0}=1584.8\,\mathrm{cm^{-1}}$, $\gamma_{0}=7.6\,\mathrm{cm^{-1}}$, $\gamma_{\mathrm{el}}=395\,\mathrm{cm^{-1}}$, $\lambda=4\times 10^{-3}$ and $v_\mathrm{F}=1.33\times 10^6\,\mathrm{m/s}$, which we then use to calculate the maximum width $\Gamma_{\mathrm{G},T_1}^\mathrm{max}$ at the resonance $B_{T_\mathrm{1}}$ as a function of charge carrier density, as represented by the blue line in Fig.~2a in the main text.

			\begin{figure}[b]
				\includegraphics{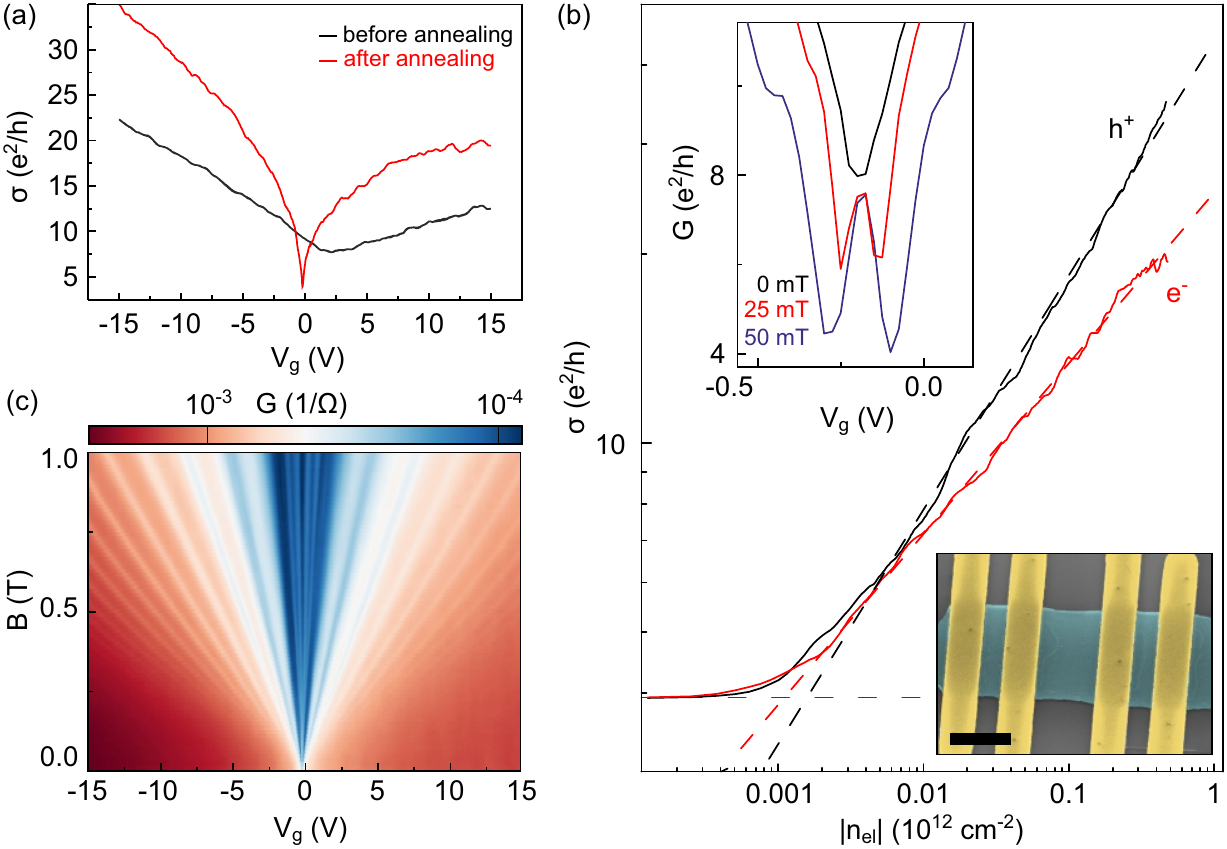}
				\caption{(a) Electrical conductivity $\sigma$  as a function of gate voltage $V_\mathrm{g}$ of the suspended graphene sample before and after current annealing. Measurements are taken at $T=4.2$~K. (b) Double logarithmic graph of the conductivity $\sigma$ after current annealing as a function of charge carrier density for holes (black) and electrons (red). The red and black dashed lines are linear fits outside the regime of charge inhomogeneity. The purple, horizontal, dashed line indicates the minimum conductivity due to charge inhomogeneity. The crossing point of these lines defines $n^*$. The left inset shows the onset of SdHO at approx. 25~mT, indicating a mobility $\mu\,{\approx}\,400\,000\,\mathrm{cm^2/(Vs)}$. The right inset depicts a false-color scanning electron micrograph of the measured device. The scale bar represents $2\,$\textmu m. (c) Electrical conductivity as a function of magnetic field and gate voltage, used to determine back gate the lever-arm $\alpha=3.15\times10^{10}\,\mathrm{cm^{-2}V^{-1}}$.}
				\label{fig:Device}
			\end{figure}
			
			\begin{figure}[tb]
				\includegraphics{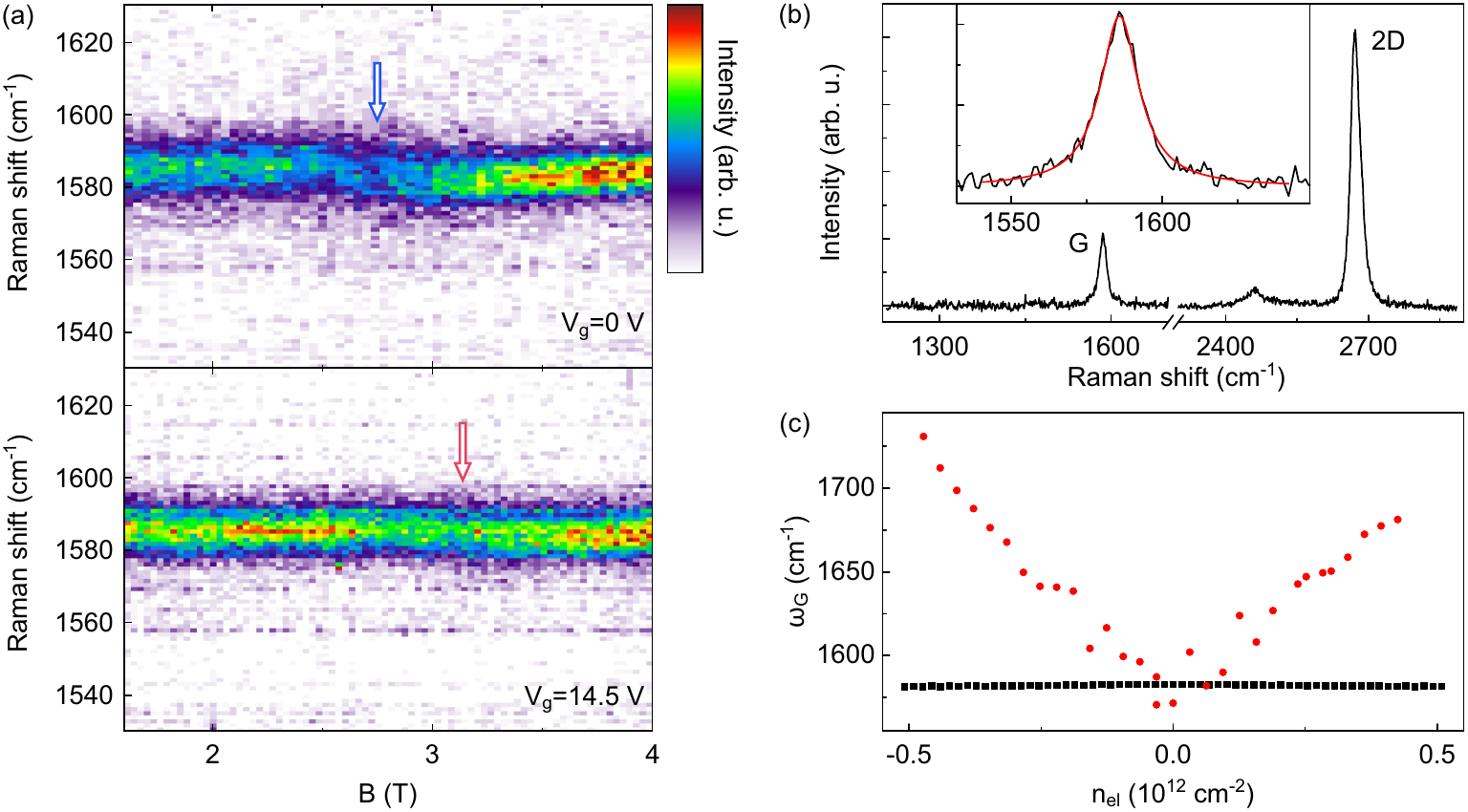}
				\caption{(a) Raman intensity around the G peak plotted as a function of the magnetic field around the $T_1$-MPR at $V_\mathrm{g}=0\,\mathrm{V}$ and $V_\mathrm{g}=14.5\,\mathrm{V}$. The arrows indicate the positions of the $T_1$-MPRs as extracted in the main text. (b) Raman spectrum taken at $V_\mathrm{g}=0\,\mathrm{V}$ and $B=0\,\mathrm{T}$. the inset depicts a Lorentzian fit of the G~peak. (c) Illustration of the change in $\omega_\mathrm{G}$ required to observe the shift in $B_\mathrm{T_1}$ as seen in Fig.~2 in the main text under the assumption of a constant $v_\mathrm{F}=1.06\times 10^6$~m/s (red dots).  Black squares represent the actually measured $\omega_\mathrm{G}$ as a function of charge carrier density at zero magentic field.}
				\label{fig:omega0}
			\end{figure}

	\section{Tight-Binding Model}
	\label{sec:TB}
	
	We use a third-nearest neighbor tight-binding description of
	graphene \cite{Chizhova2015} to evaluate the Coulomb and exchange
	contributions required for the self-energy correction
	$\Sigma_{\pm n}^{\mathrm{HF}}$.
	We include the magnetic field via a Peierls
	phase factor and eliminate edge states by a finite mass boundary term
	at the zigzag edges \cite{Chizhova2015}.
	We use an Arnoldy-Lanczos algorithm
	in conjunction with shift-invert \cite{Lanczos1950} to calculate approximately 3000 eigenstates (in groups of 400 for
	efficiency) of a quadratic graphene flake of size $40\times 40
	\;\mathrm{nm}^2$ at a magnetic field of 200 T.
	The Coulomb and exchange contributions are evaluated as
	\begin{equation}
		v^{\mathrm{Hart.}}_{i,j} = e^2 \bra{ij}\frac{1}{\left|\vec{r}_1-\vec{r}_2\right|}\ket{ij}
	\end{equation}
	and
	\begin{equation}
		v^{\mathrm{Fock}}_{i,j} = e^2 \bra{ji}\frac{1}{\left|\vec{r}_1-\vec{r}_2\right|}\ket{ij}.
	\end{equation}
	Considering the scaling invariance of the Dirac equation, we
	rescale our results down to the experimental field strength
	($\l_B \propto 1/\sqrt{B}$).
	Effectively, our results thus
	correspond to an $\approx 330\times 330$ nm$^2$-sized flake. We evaluate both
	contributions for all pairs of eigenstates.
	The state indices $i$ and $j$ can each be split into a Landau level index $\pm n$ and a quantum number $m$, that labels the degenerate states, as described in the main manuscript, i.e., $i = (\pm n,m)$, $j = (\pm n',m')$.
	Eigenstates are assigned to specific Landau levels $\pm n$ and $\pm n'$
	based on their energy.
	Due to the finite size of our system, there are a few states within the energy gaps between the Landau levels, which we do not include in the evaluation of the self energy.

	\section{Temperature-dependent Shubnikov-de-Haas oscillations}
	
\begin{figure}[tb]
		\includegraphics[width=\columnwidth]{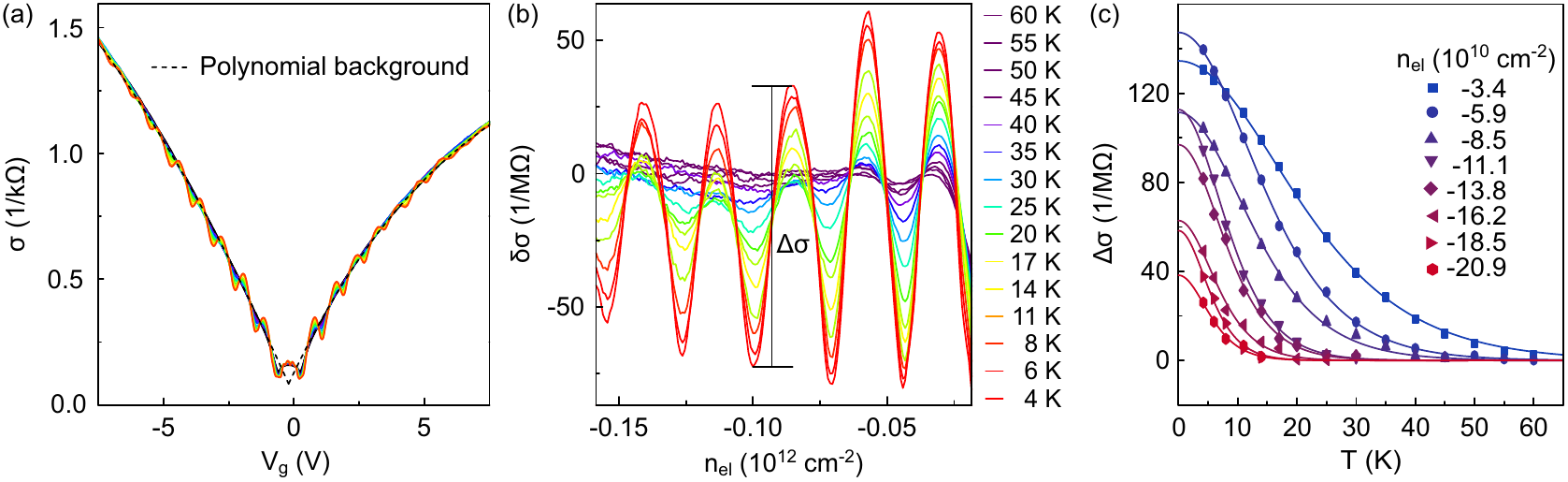}
		\caption{(a) Electrical conductivity as a function of applied gate voltage $V_\mathrm{g}$ for different temperatures at a magnetic field of $B=0.25$~T. The dashed line represent a fit of a polynomial background for both electron and hole doping. (b) Shubnikov-de Haas oscillations at  $B=0.25$~T after subtraction of a polynomial background. (c) Height of the oscillations $\Delta\sigma$ as a function of temperature for different charge carrier densities.}
		\label{fig:SdHO}
	\end{figure}

	To extract the renormalization of $v_\mathrm{F}$ at low magnetic fields, we closely follow the method used by Elias~\textit{et al.} \cite{Elias2011} and perform temperature-dependent Shubnikov-de Haas oscillation measurements.
	We measure the conductivity $\sigma$ as a function of charge carrier density for different temperatures in a range of $T=4$~K to $T=60$~K at a magnetic field of $B=0.25$~T (see Fig.~\ref{fig:SdHO}a). %
	For electrons and holes we separately fit a 4th-order polynomial to the smooth, high temperature data and subtract this background from all measurements.
	The resulting conductivity oscillations $\delta\sigma$ are shown exemplary in Fig.~\ref{fig:SdHO}b for the hole side. We extract the amplitude of the Shubnikov-de Haas oscillations as the difference $\Delta\sigma$ between maxima and minima (see label in Fig.~\ref{fig:SdHO}b). This makes the extracted amplitude almost independent of the chosen background.
	The amplitude as a function of temperature ($T$) for different hole densities are presented in Fig.~\ref{fig:SdHO}c. The amplitude follow the Lifshitz-Kosevich formula \cite{Sharapov2004}:
	\begin{equation}
		\Delta \sigma \propto T/\mathrm{sinh}\left(2\pi^2k_\mathrm{B}T \frac{m_\mathrm{c}}{e\hbar B}\right),
	\end{equation}
	where $m_c=\hbar\sqrt{\pi n_\mathrm{el}}/v_\mathrm{F}$ is the cyclotron mass at a given $n_\mathrm{el}$. By fitting this expression to our data (see Fig.~\ref{fig:SdHO}c), we are able to extract $v_\mathrm{F}$ for different charge carrier densities, which is shown in the inset in Fig.~3 in the main text.
	\label{sec:SdHO}

	\section{Finite renormalization of the Fermi velocity in presence of Landau levels}
	
	As shown by Gonz\'{a}lez \textit{et al.} \cite{Gonzalez1994}, for two-dimensional massless Dirac fermions interacting via the Coulomb potential, the Fock contribution to the Fermi velocity is logarithmically divergent in the limit of zero temperature and chemical potential.
	The corresponding Fock contribution to the Hamiltonian is:
	\begin{equation}
		\hat{H}_F=\sum_{\vec{k}}\sum_{\alpha,\beta}\hat{\Psi}_\alpha^\dagger(\vec{k})h_{\alpha\beta}(\vec{k})\hat{\Psi}_\beta(\vec{k}),
	\end{equation}
	where $\hat{\Psi}_\beta(\vec{k})$ is the annihilation operator for the Dirac fermion with the quasi-wave vector $\vec{k}$ and pseudospin projection $\beta$,
	\begin{equation}
		h_{\alpha\beta}(\vec{k})=-2\pi e^2\sum_{\vec{k}'}\frac{\rho_{\beta\alpha}(\vec{k'})}{|\vec{k}-\vec{k}'|},
	\end{equation}
	and $\rho_{\beta\alpha}(\vec{k})=\braket{\hat{\Psi}^\dagger_{\vec{k}\alpha}\hat{\Psi}^{\,}_{\vec{k}\beta}}$ is the single-particle density matrix (\cite{Katsnelson2012}, Sect. 8.4).
	Its spinor structure is given by the expression:
	\begin{equation}\label{eqn:rho}
		\hat{\rho}_{\vec{k}}=n_{\vec{k}}\hat{I}+\vec{m}_{\vec{k}}\hat{\vec{\sigma}},
	\end{equation}
	where $\hat{I}$ is the $2\times2$ unit matrix, $\hat{\vec{\sigma}}$ are the Pauli matrices and the “pseudospin density” $\vec{m}_{\vec{k}}$ has the form $\vec{m}_{\vec{k}}=\vec{k}F(k)$.
	Following Ref.~\cite{Katsnelson2012}, Sect. 7.2, for chemical potential and temperature equal to zero
	\begin{equation}\label{eqn:F}
		F(k)=\frac{1}{2k}.
	\end{equation}
	Then the Fock contribution to the Fermi velocity reads (\cite{Katsnelson2012}, Sect. 8.4):
	\begin{equation}\label{eq:vFwoB}
		\delta v_\mathrm{F} =\frac{\pi e^2}{\hbar}\sum_{\vec{k}}\frac{F(k)}{k}=\frac{e^2}{2\hbar}\int_{0}^{\Lambda}\mathrm{d}k\,F(k),
	\end{equation}
	where $\Lambda\propto1/a$ is the ultraviolet (UV) cutoff due to the inapplicability of the Dirac model at large wave vectors and $a$ is the interatomic distance of the graphene lattice.
	Explicit numerical calculations on a lattice for the case of a pure Coulomb interaction \cite{Astrakhantsev2015} give the value $\Lambda\approx 0.8/a$. When substituting Eq.~\ref{eqn:F} into Eq.~\ref{eq:vFwoB} we have a divergence at the lower limit, which, at finite charge carrier density, is cut off at the Fermi wave vector $k_\mathrm{F}$. The result reads:
	\begin{equation}\label{eq:vFlog}
		\delta v_\mathrm{F}=\frac{e^2}{4\hbar}\mathrm{ln}\left(\frac{\Lambda}{k_\mathrm{F}} \right).
	\end{equation}
	
	In the presence of a magnetic field, the density matrix Eq.~\ref{eqn:rho} and therefore the function $F(k)$ can be calculated using the explicit expression for the Green's function of massless Dirac electrons in the presence of a magnetic field found in Ref.~\cite{Gusynin1995}. The result is
	\begin{equation}\label{eq:FinB}
		F(k)=\frac{l_B}{2\sqrt{\pi}}\int_{0}^{\infty}\mathrm{d}s\,\frac{\exp(-k^2 l_B^2 \tanh(s))}{\sqrt{s}\cosh(s)^2},
	\end{equation}
	where $l_B=\sqrt{\hbar/(eB)}$ is the magnetic length. Substituting Eq.~\ref{eq:FinB} into Eq.~\ref{eq:vFwoB} and changing the order of integrations we obtain
	\begin{equation}\label{eq:vFinB}
		\delta v_\mathrm{F}=\frac{e^2}{8\hbar}\int_{0}^{\infty}\mathrm{d}s\,\frac{\mathrm{erf}(\Lambda l_B \sqrt{\tanh(s)})}{\sqrt{s \tanh(s)}\cosh(s)^2},
	\end{equation}
	where
	\begin{equation}
		\mathrm{erf}(x)=\frac{2}{\sqrt{\pi}}\int_{0}^{x}\mathrm{d}s\,\exp(-x^2)
	\end{equation}
	is the error function. Assuming that $\Lambda l_B\gg1$,  one can calculate the integral in Eq.~\ref{eq:vFinB} by splitting the integration interval into two parts: $(0,\infty)=(0,C)+(C,\infty)$ with some ${1/(\Lambda l_B)^2\ll C \ll 1}$. With logarithmic accuracy, one has, instead of Eq.~\ref{eq:vFlog},
	\begin{equation}
		\delta v_\mathrm{F}=\frac{e^2}{4\hbar}\mathrm{ln}(\Lambda l_B).
	\end{equation}
	Thus the infrared divergence (Eq.~\ref{eq:vFlog}) is cut off at wave vectors on the order of the inverse magnetic length and the dependence of the Fermi velocity on the electron filling factor is no longer singular in the presence of a magnetic field.


\begin{thebibliography}{45}
\bibitem [{sup()}]{supp}%
  \BibitemOpen
  \href@noop {} {}\bibinfo {howpublished} {See Supplemetal Material [URL will
  be inserted by publisher] for additional details on the tight-binding
  calculations, the model used to fit the MPRs and more detailed electrical
  characterisations.}\BibitemShut {Stop}%
\end{thebibliography}
\end{document}